\documentstyle[preprint,aps,12pt]{revtex}
\begin{document}
\draft

\preprint{YITP-97-52, \ hep-ph/9710515}
\title{Non-factorizable long distance contributions \\
in color suppressed decays of $B$ mesons
\footnote{Revised edition of YITP-97-26}
} 
\author{ K. Terasaki}

\address{Yukawa Institute for Theoretical Physics, Kyoto University,
Kyoto 606-01, Japan}
\date{ October, 1997}
\maketitle

\begin{abstract} 
$\bar B \rightarrow D\pi$, $D^*\pi$, $J/\psi\bar K$ and $J/\psi\pi$ 
decays are studied. Their amplitude is given by a sum of factorized 
and non-factorizable ones. The latter which is estimated by using a 
hard pion approximation is rather small in color favored 
$\bar B \rightarrow D\pi$ and $D^*\pi$ decays but still can
efficiently interfere with the main amplitude given by the
factorization. In the color suppressed 
$\bar B \rightarrow J/\psi\bar K$ and $J/\psi\pi$ decays, the 
non-factorizable contribution is very important. The sum of the 
factorized and non-factorizable amplitudes can reproduce well the 
existing experimental data on the branching ratios for the color 
favored $\bar B \rightarrow D\pi$ and $D^*\pi$ and the color 
suppressed $\bar B \rightarrow J/\psi \bar K$ and $J/\psi\pi$ decays 
by taking reasonable values of unknown parameters involved. 

\end{abstract}
 
\vskip 1cm
\newpage

\section{Introduction}

Nonleptonic weak decays of charm and $B$ mesons have been studied 
extensively\cite{BSW,NRSX} by using the so-called factorization 
(or vacuum insertion) prescription. It has been supported by two 
different arguments. One is that the factorization is applicable in 
the large $N_c$ (color degree of freedom) limit\cite{Large-N} and the 
other is that it can be a good approximation under a certain 
kinematical condition\cite{DG}, {\it i.e.}, a heavy quark decays into 
another heavy quark plus a pair of light quark and anti-quark which
are emitted colinearly with sufficiently high energies, for example, 
like $b \rightarrow c\,+\, (\bar ud)_1$, where $(\bar ud)_1$ denotes a 
color singlet $(\bar ud)$ pair. 

The factorization predicts the so-called color suppression
[suppression of color mismatched decays like 
$\bar B^0 \rightarrow D^0\pi^0$, $D^{*0}\pi^0$, etc., described by 
$b \rightarrow (c\bar u)_8 \,+\,d$, where $(c\bar u)_8$ means that the 
$(c\bar u)$ pair is of color octet]. According to recent measurements 
of $\bar B \rightarrow D\pi$ decays\cite{BHP}, 
the $\bar B^0 \rightarrow D^0\pi^0$ decay is actually suppressed in 
comparison with color favored decays like $B^- \rightarrow D^0\pi^-$, 
etc., described by $b \rightarrow c\,+\,(\bar ud)_1$. However, 
semi-phenomenological analyses\cite{CLEO,Kamal} in two-body decays of 
$B$ mesons within the framework of the factorization suggest that the 
value of $a_2$ to reproduce the observed branching ratios for these
decays\cite{CLEO,PDG} should be larger by about a factor 2 than the 
one with the leading order (LO) QCD corrections\cite{BSW,NRSX,HW} 
where $N_c = 3$ and that its sign should be opposite to the one in 
the large $N_c$ limit although the phenomenological value of $a_1$ 
is very close to the one expected in the same approximation. 
[$a_1$ and $a_2$ are the coefficients of four quark operators in the 
effective weak Hamiltonian in the Bauer-Stech-Wirbel (BSW) 
scheme\cite{BSW,NRSX} which will be reviewed briefly in the next 
section.] The above fact implies that the large $N_c$ argument fails, 
at least, in hadronic weak decays of $B$ mesons. Since the large 
$N_c$ argument is independent of flavors, it also does not work in 
nonleptonic weak decays of charm mesons. The kinematics of charm 
meson decays is far from the condition, {\it i.e.}, like 
$b \rightarrow c\, +\, (\bar ud)_1$ mentioned before, under which the 
factorization is applicable. Therefore the factorization of charm 
decay amplitudes has no theoretical support. In fact, a naive 
application of the factorization to charm decay amplitudes again 
leads to the color suppression [suppression of color mismatched 
decays, $D^0 \rightarrow \bar K^0\pi^0,\,\bar K^{*0}\pi^0$, etc.,
described by $c \rightarrow (s\bar d)_8 \,+\, u$] and therefore the 
amplitudes for two body decays of charm mesons must be approximately 
real. However the observed decay rates for these decays are not 
always suppressed and the amplitudes for $D \rightarrow \bar K\pi$ 
and $\bar K^*\pi$ decays have large phases\cite{BHP}. In this way, 
it will be understood that the factorized amplitudes might be 
dominant only in some specific decays like the color favored 
$\bar B \rightarrow D\pi$ and $D^*\pi$ decays\cite{DG}. The above 
result on the $B$ decays that the phenomenological value of $a_2$ is 
much larger than the one with the LO corrections seems to suggest
that, even in the decays of $B$ mesons with high mass, factorization 
is not complete but non-factorizable long distance hadron dynamics 
still cannot be neglected. Buras\cite{BURAS} calculated 
next-to-leading order (NLO) QCD corrections to the effective weak 
Hamiltonian in three different renormalization schemes and observed 
that $a_2$ can have large NLO corrections which are strongly 
dependent on choice of renormalization scheme while the corrections 
to $a_1$ are very small and are stable with respect to change of 
renormalization scheme. Then he discussed that the above instability 
of the NLO corrections to $a_2$ may imply importance of 
non-factorizable effects on $B$ decays. Soares\cite{SOARES} tried to 
estimate phenomenologically non-factorizable contributions to 
$\bar B \rightarrow D\pi$ and $\bar B \rightarrow J/\psi\bar K$ 
decays and found several possible solutions which indicate large 
non-factorizable contributions. 

In this article, we study $\bar B \rightarrow D\pi$, $D^*\pi$, 
$J/\psi\bar K$ and $J/\psi\pi$ decays describing the amplitude for
these decays by a sum of factorizable and non-factorizable ones. 
The latter amplitude is estimated by using a hard pion (or kaon) 
approximation. The $\bar B \rightarrow D\pi$ and $D^*\pi$ decays will 
be studied in the next section. It will be seen that, in the color 
suppressed $\bar B^0 \rightarrow D^0\pi^0$ and $D^{*0}\pi^0$ decays, 
the hard pion amplitudes as the non-factorizable long distance 
contributions are important. In the section 3, the decays, 
$\bar B \rightarrow J/\psi\bar K$ and $B^- \rightarrow J/\psi\pi^-$, 
both of which are color suppressed, will be investigated in the same 
way. It will be demonstrated that, in these decays, non-factorizable 
long distance amplitudes are again important. 
A brief summary will be given in the final section. 

\section{$\bar B \rightarrow D\pi$ and $D^*\pi$ decays}

Our starting point is to describe the amplitude for two body decay of 
$B$ meson by a sum of factorizable and non-factorizable
ones\cite{BDP}, 
\begin{equation}
M_{\rm total} = M_{\rm fact} + M_{\rm non-f}.          \label{eq:AMP}
\end{equation}
The factorizable amplitude $M_{\rm fact}$ is evaluated by using 
the factorization in the BSW scheme\cite{BSW,NRSX} in which the 
relevant part of the effective weak Hamiltonian is given by 
\begin{equation}
H_w^{BSW} 
= {G_F\over\sqrt{2}}U_{cb}U_{ud}
                   \Bigl\{a_1O_1^H + a_2O_2^H + h.c.\Bigr\}.    
                                                      \label{eq:BSW}
\end{equation}
It can be obtained by applying the Fierz reordering to the 
usual effective Hamiltonian 
\begin{equation}
H_w^{} 
= {G_F\over\sqrt{2}}U_{cb}U_{ud}
                        \Bigl\{c_1O_1 + c_2O_2 + h.c.\Bigr\},
                                                      \label{eq:WH}
\end{equation}
where $c_1$ and $c_2$ are the Wilson coefficients of the four quark
operators,   
\begin{equation}
O_1 = [(\bar du)_L\,+\,(\bar sc)_L](\bar cb)_L,  \qquad
O_2 = (\bar cu)_L(\bar db)_L \,+\,(\bar cc)_L(\bar sb)_L
                                                       \label{eq:FQO}
\end{equation}
with $(\bar q'q)_L=\bar q'\gamma_\mu (1 - \gamma_5)q$. 
The quark bilinears in $O_1^H$ and $O_2^H$ are treated as
interpolating fields for the mesons and therefore should be no longer 
Fierz reordered. 
The coefficients, $a_1$ and $a_2$, in Eq.(\ref{eq:BSW}) are given by 
\begin{equation}
a_1 = c_1 + {c_2 \over N_c}, \qquad a_2 = c_2 + {c_1 \over N_c}.
                                                     \label{eq:WC}
\end{equation}
The LO QCD corrections lead to $a_1\simeq 1.03$ and $a_2\simeq 0.11$ 
for $N_c=3$\cite{NRSX}. $U_{ij}$ is the Cabibbo-Kobayashi-Maskawa 
(CKM) matrix element\cite{CKM} which is taken to be real in this 
article since $CP$ invariance is always assumed. 

The factorization prescription in the BSW scheme leads to the
following factorized amplitude, for example, for the 
$B^-(p) \rightarrow D^0(p')\pi^-(q)$ decay, 
\begin{eqnarray}
 M_{\rm fact}(B^-(p) \rightarrow D^0(p')\pi^-(q))
= {G_F \over \sqrt{2}}U_{cb}U_{ud}
\Bigl\{&& 
a_1\langle \pi^-(q)|(\bar du)_L|0\rangle 
\langle D^0(p')|(\bar cb)_L|B^-(p) \rangle         \nonumber\\
&& + a_2\langle D^0(p')|(\bar cu)_L|0\rangle
\langle \pi^-(q)|(\bar db)_L|B^-(p) \rangle
\Bigr\}.
                                                    \label{eq:FACT}
\end{eqnarray}
\newpage
\begin{center}
\begin{quote}
{Table~I. Factorized amplitudes for $\bar B \rightarrow D\pi$ and 
$D^*\pi$ decays ($m_\pi^2=0$).}
\end{quote}
\vspace{0.5cm}

\begin{tabular}
{l|l}
\hline
$\quad\,\,${\rm Decay}
&\hskip 5cm {$\quad M_{\rm fact}\,$}
\\
\hline 
$\bar B^0 \rightarrow D^{+}\pi^-$
& $\,\quad iU_{cb}U_{ud}
   {G_F \over\sqrt{2}}a_1f_\pi(m_B^2 - m_D^2)F_0^{DB}(0) 
\Bigl[
1 - \Bigl({a_2 \over a_1}\Bigr)\Bigl({f_B \over f_\pi}\Bigr)
\Bigl({m_D^2 \over m_B^2 - m_D^2}\Bigr)
{F_0^{D\pi}(m_B^2) \over F_0^{DB}(0)}
\Bigr]$ 
\\
$\bar B^0 \rightarrow D^{0}\pi^0$
& $\,-\,iU_{cb}U_{ud}{G_F \over {2}}a_2f_Dm_B^2F_0^{\pi B}(m_D^2) 
\Bigl[
1 + \Bigl({f_B \over f_D}\Bigr)
\Bigl({m_D^2 \over m_B^2}\Bigr)
{F_0^{D\pi}(m_B^2) \over F_0^{\pi B}(m_D^2)}
\Bigr]
$
\\
$B^- \rightarrow D^0\pi^-$
& $\,\quad iU_{cb}U_{ud}{G_F \over\sqrt{2}}a_1f_\pi(m_B^2 - m_D^2)F_0^{DB}(0) 
\Bigl[
1 + \Bigl({a_2 \over a_1}\Bigr)\Bigl({f_D \over f_\pi}\Bigr)
\Bigl({m_B^2 \over m_B^2 - m_D^2}\Bigr)
{F_0^{\pi B}(m_D^2) \over F_0^{DB}(0)}
\Bigr]
$ 
\\
$\bar B^0 \rightarrow D^{*+}\pi^-$
& $\,-\,iU_{cb}U_{ud}{G_F \over\sqrt{2}}
a_1f_\pi A_0^{D^*B}(0) 
\Bigl[
1 - \Bigl({a_2 \over a_1}\Bigr)\Bigl({f_B \over f_\pi}\Bigr)
{A_0^{D^*\pi}(m_B^2) \over A_0^{D^*B}(0)}
\Bigr]2m_{D^*}\epsilon^*(p')\cdot p$
\\
$\bar B^0 \rightarrow D^{*0}\pi^0$
& $\,\quad iU_{cb}U_{ud}{G_F \over {2}}a_2f_{D^*}
F_1^{\pi B}(m_D^{*2}) 
\Bigl[
1 + \Bigl({f_B \over f_{D^*}}\Bigr)
{A_0^{D^*\pi}(m_B^2) \over F_1^{\pi B}(m_D^{*2})}
\Bigr]2m_{D^*}\epsilon^*(p')\cdot p$
\\
$B^- \rightarrow D^{*0}\pi^-$
& $\,-\,iU_{cb}U_{ud}{G_F \over\sqrt{2}}
a_1f_\pi A_0^{D^*B}(0) 
\Bigl[
1 + \Bigl({a_2 \over a_1}\Bigr)\Bigl({f_{D^*} \over f_\pi}\Bigr)
{F_1^{\pi B}(m_D^{*2}) \over A_0^{D^*B}(0)}
\Bigr]2m_{D^*}\epsilon^*(p')\cdot p$ 
\\
\hline
\end{tabular}

\end{center}
\vspace{0.5cm}
Factorizable amplitudes for the other $\bar B \rightarrow D\pi$ and
$D^*\pi$ decays also can be calculated in the same way. 
To evaluate these amplitudes, we use the parameterization of matrix 
elements of currents in Ref.\cite{NRSX},  
\begin{equation}
       \langle \pi(q)|A_\mu^{}|0 \rangle = -if_{\pi}q_\mu, 
\qquad \langle 0|A_\mu^{}|\pi(q) \rangle = if_{\pi}q_\mu,   
                                                       \label{eq:PCAC}
\end{equation}
\begin{eqnarray}
&&\langle D(p')|V_\mu^{}|\bar B(p)\rangle 
= \Biggl\{(p+p')_\mu - {m_B^2 - m_D^2 \over q^2}q_\mu\Biggr\}F_1(q^2) 
          + {m_B^2 - m_D^2 \over q^2}q_\mu F_0(q^2),  \label{eq:DB} \\
&&\langle D^*(p')|A_\mu^{}|\bar B(p)\rangle 
= \Biggl\{
(m_B + m_{D^*})\epsilon_\mu^*(p')A_1(q^2) 
- {\epsilon^*(p')\cdot q \over m_B + m_{D^*}}(p + p')_\mu A_2(q^2)
\nonumber \\
&& \hskip 6cm
- 2m_{D^*}{\epsilon^*\cdot q \over q^2}q_\mu A_3(q^2)
\Biggr\}
+ 2m_{D^*}{\epsilon^*\cdot q \over q^2}q_\mu A_0(q^2), 
                                                     \label{eq:D^*B}
\end{eqnarray}
where $q = p - p'$ and the form factors satisfy 
\begin{eqnarray}
&& A_3(q^2) = {m_B + m_{D^*} \over 2m_{D^*}}A_1(q^2) 
- {m_B - m_{D^*} \over 2m_{D^*}}A_2(q^2),           \label{eq:FFA}\\
&& 
F_1(0) = F_0(0), \qquad A_3(0) = A_0(0).           \label{eq:FF0}
\end{eqnarray}
To get rid of useless imaginary unit except for the overall phase in 
the amplitude, however, we adopt the following parameterization of 
matrix element of vector current\cite{HY} 
\begin{equation}
\langle V(p')|V_\mu^{}|0\rangle 
        = -if_{V}m_{V}\epsilon_{\mu}^*(p').      \label{eq:Lvector}
\end{equation}
As stressed in Ref.\cite{HY}, the above matrix element of vector 
current can be treated in parallel to those of axial vector currents 
in Eq.(\ref{eq:PCAC}) in the infinite momentum frame (IMF). 
Using these expressions of current matrix elements, we obtain 
the factorized amplitudes for $\bar B \rightarrow D\pi$ and 
$D^*\pi$ decays in Table~I, where we have put $m_\pi^2=0$. 

Before we evaluate numerically the factorized amplitudes, we 
study non-factorizale amplitudes for $\bar B \rightarrow D\pi$ and 
$D^*\pi$ decays using a hard pion technique in the IMF; 
${\bf p} \rightarrow \infty$\cite{HARDP,suppl}. 
In our hard pion approximation, the non-factorizable amplitude 
for the $\bar B({p}) \rightarrow D({p'})\pi({q})$ decay is given by 
\begin{equation}
M_{\rm non-f}(\bar B \rightarrow D\pi) 
\simeq M_{\rm ETC}(\bar B \rightarrow D\pi) 
+ M_{\rm S}(\bar B \rightarrow D\pi).        
                                                      \label{eq:HP}
\end{equation} 
The equal-time commutator term ($M_{ETC}$) and the surface term
($M_S$) are given by 
\begin{equation}
M_{\rm ETC}(\bar B \rightarrow D\pi) 
= {i \over f_{\pi}}\langle{D|[V_{\bar \pi}, H_w]|\bar B}\rangle  
                                                    \label{eq:ETC}
\end{equation}
and
\begin{eqnarray} 
M_{\rm S}(\bar B \rightarrow D\pi) &&                       
= {i \over f_{\pi}}\Biggl\{\sum_n\Bigl({m_D^2 - m_B^2 
                                         \over m_n^2 - m_B^2}\Bigr)
  \langle{D|A_{\bar \pi}|n}\rangle
                         \langle{n|H_w|\bar B}\rangle  \nonumber \\
&& \qquad\qquad 
+ \sum_\ell\Bigl({m_D^2 - m_B^2 
                              \over m_\ell^2 - m_D^2}\Bigr)
\langle{D|H_w|\ell}\rangle
                  \langle{\ell|A_{\bar \pi}|\bar B}\rangle\Biggr\},  
                                                    \label{eq:SURF}
\end{eqnarray}
respectively, where $[V_\pi + A_\pi, H_w]=0$ has been used. (See 
Refs.\cite{HARDP} and \cite{suppl} for notations.) $M_{ETC}$ and 
$M_S$ have to be evaluated in the IMF. The surface term has been 
given by a sum of all possible pole amplitudes, {\it i.e.}, $n$ and 
$\ell$ run over all possible single mesons, not only ordinary 
$\{q\bar q\}$, but also hybrid $\{q\bar qg\}$ and exotic 
$\{qq\bar q\bar q\}$ mesons. Since the $B$ meson mass $m_B$ is much 
higher than those of charm mesons and since wave function 
overlappings between the ground-state $\{q\bar q\}_0$ and 
excited-state-meson states are expected to be small, however, excited 
meson contributions will be small in these decays and can be safely 
neglected. Therefore the hard pion amplitudes as the non-factorizable 
long distance contributions are approximately described in terms of 
{\it asymptotic ground-state-meson matrix elements} (matrix elements 
taken between single ground-state-meson states with infinite 
momentum) of $V_\pi$, $A_\pi$ and $H_w$. Hard pion amplitudes for 
$\bar B \rightarrow D^*\pi$ decays can be obtained by exchanging $D$ 
for $D^*$.

Asymptotic matrix elements of $V_{\pi}$ and $A_{\pi}$  are
parameterized as 
\begin{eqnarray}
&&\langle{\pi^0|V_{\pi^+}|\pi^-}\rangle 
    = \sqrt{2}\langle{K^{+}|V_{\pi^+}|K^0}\rangle 
    = -\sqrt{2}\langle{D^{+}|V_{\pi^+}|D^0}\rangle 
    =  \sqrt{2}\langle{B^{+}|V_{\pi^+}|B^0}\rangle 
    = \cdots = \sqrt{2},                            \label{eq:MEV}\\
&&\langle{\rho^0|A_{\pi^+}|\pi^-}\rangle 
    = \sqrt{2}\langle{K^{*+}|A_{\pi^+}|K^0}\rangle 
    = -\sqrt{2}\langle{D^{*+}|A_{\pi^+}|D^0}\rangle 
    =  \sqrt{2}\langle{B^{*+}|A_{\pi^+}|B^0}\rangle 
    = \cdots = h,                                    \label{eq:MEA}
\end{eqnarray}
where $V_\pi$'s and $A_\pi$'s are isospin charges and their axial
counterpart, respectively. The above parameterization can be obtained 
by using asymptotic $SU_f(5)$ symmetry\cite{ASYMP}, or $SU_f(5)$ 
extension of the nonet symmetry in $SU_f(3)$. Asymptotic matrix 
elements of $V_\pi$ between vector meson states can be obtained by 
exchanging pseudo scalar mesons for vector mesons with corresponding 
flavors in Eq.(\ref{eq:MEV}), for example, as 
$\pi^{0,-} \rightarrow \rho^{0,-}$, etc. 
The $SU_f(4)$ part of the above parameterization reproduces 
well\cite{suppl,Takasugi} the observed values of decay rates, 
$\Gamma(D^* \rightarrow D\pi)$ and $\Gamma(D^* \rightarrow D\gamma)$. 

Amplitudes for dynamical hadronic processes can be decomposed into 
({\it continuum contribution}) + ({\it  Born term}).  
Since $M_{\rm S}$ is given by a sum of pole amplitudes, $M_{\rm ETC}$ 
corresponds to the continuum contribution \cite{MATHUR} which can 
develop a phase relative to the Born term. Therefore we parameterize
the ETC terms using isospin eigen amplitudes and their phases. Since 
the $D\pi$ final states can have isospin $I={1 \over 2}$ and 
${3 \over 2}$, we decompose $M_{\rm ETC}$'s as  
\begin{eqnarray}
&&M_{\rm ETC}(\bar B^0\, \rightarrow D^+\pi^-) 
= \,\,\,\,\sqrt{{1 \over 3}}M_{\rm ETC}^{(3)}e^{i\delta_{3}} 
+ \sqrt{{2 \over 3}}M_{\rm ETC}^{(1)}e^{i\delta_{1}},  \label{eq:ETCMP}\\
&&M_{\rm ETC}(\bar B^0\, \rightarrow D^0\,\pi^0\,) 
= -\sqrt{{2 \over 3}}M_{\rm ETC}^{(3)}e^{i\delta_{3}} 
+  \sqrt{{1 \over 3}}M_{\rm ETC}^{(1)}e^{i\delta_{1}},  
                                                \label{eq:ETCZZ}\\
&&M_{\rm ETC}(B^- \rightarrow  D^0\pi^-) 
= \quad\sqrt{3}M_{\rm ETC}^{(3)}e^{i\delta_{3}},            
                                                \label{eq:ETCZP}
\end{eqnarray}
where $M_{\rm ETC}^{(2I)}$'s are the isospin eigen amplitudes with
isospin $I$ and $\delta_{2I}$'s are the corresponding phase shifts
introduced. In the present approach, therefore, the final state 
interactions are included in the non-factorizable amplitudes. 

In this way we can describe the non-factorizable amplitudes for the 
$\bar B \rightarrow D\pi$ decays as 
\begin{eqnarray}
&&M_{\rm non-f}(\bar B^0 \rightarrow D^+\pi^-) \nonumber\\
&&\qquad
\simeq -i{\langle{D^0|H_w|\bar B^0}\rangle \over f_\pi}
\Biggl\{
\Biggl[{4\over 3}e^{i\delta_1} -{1\over 3}e^{i\delta_3}\Biggr]
+ {\langle{D^{*0}|H_w|\bar B^0}\rangle \over 
                            \langle{D^0|H_w|\bar B^0}\rangle}
\Biggl({m_{B}^2-m_D^2 \over m_{B}^2-m_{D^*}^2}\Biggr)               
          \sqrt{1\over 2}h + \cdots
                                     \Biggr\},  \label{eq:LMP}
\end{eqnarray}
\newpage
\begin{eqnarray}
&&M_{\rm non-f}(\bar B^0 \rightarrow D^0\pi^0) 
\simeq -i{\langle{D^0|H_w|\bar B^0}\rangle \over f_\pi}
\Biggl\{
{\sqrt{2} \over 3}\Biggl[2e^{i\delta_1} + e^{i\delta_3}\Biggr]
\nonumber\\
&&\qquad
+ \sqrt{1 \over 2}\Biggl[{\langle{D^{*0}|H_w|\bar B^0}\rangle 
                       \over \langle{D^0|H_w|\bar B^0}\rangle}
   \Biggl({m_{B}^2-m_D^2 \over m_{B}^2-m_{D^*}^2}\Biggr) 
+ {\langle{D^{0}|H_w|\bar B^{*0}}\rangle \over 
                             \langle{D^{0}|H_w|\bar B^0}\rangle}
\Biggl({m_{B}^2-m_D^2 \over m_{B^*}^2-m_D^2}\Biggr)\Biggr]
              \sqrt{1\over 2}h +\cdots
                                    \Biggr\},  \label{eq:LZZ}\\
\nonumber\\
&&M_{\rm non-f}(B^- \rightarrow D^0\pi^-) \nonumber\\
&&\qquad\qquad\qquad
\simeq i{\langle{D^{0}|H_w|\bar B^0}\rangle \over f_\pi}
\Biggl\{e^{i\delta_3} 
+ {\langle{D^{0}|H_w|\bar B^{*0}}\rangle \over 
                             \langle{D^0|H_w|\bar B^0}\rangle}
\Biggl({m_{B}^2-m_D^2 \over m_{B^*}^2-m_D^2}\Biggr)
           \sqrt{1\over 2}h +\cdots\Biggr\},   \label{eq:LZP}
\end{eqnarray}
where the ellipses denote the neglected excited meson contributions. 
The corresponding amplitudes for the $\bar B \rightarrow D^*\pi$
decays can be obtained by replacing $D^0\leftrightarrow D^{*0}$ and 
$\delta_{2I}\rightarrow\delta_{2I}^*$ in 
Eqs. (\ref{eq:LMP}) -- (\ref{eq:LZP}). Therefore the non-factorizable 
amplitudes in the hard pion approximation are controlled by the 
asymptotic ground-state-meson matrix elements of $H_w$ and the 
phases. 

Now we evaluate the amplitudes given above. The factorized amplitudes 
in Table~I contain many parameters which have not been measured by 
experiments, {\it i.e.}, form factors, $F_0(q^2)$, $A_0(q^2)$, 
$F_1(q^2)$, decay constants, $f_D$, $f_D^*$, $f_B$, etc. 
The form factors $F_0^{DB}(0)$ and $A_0^{D^*B}(0)$ can be calculated
by using the heavy quark effective theory (HQET)\cite{HQET} but the
other form factors are concerned with light meson states and 
therefore have to be estimated by some other models. In color favored
decays, main parts of the factorized amplitudes depend on the form 
factor, $F_0^{DB}(0)$ or $A_0^{D^*B}(0)$, and the other form factors 
are included in minor terms proportional to $a_2$. Therefore our 
result may not be lead to serious uncertainties although we here take 
specific values of the form factors given in Ref.\cite{Kamal}. 
In the color suppressed $\bar B^0\rightarrow D\pi^0$ and 
$D^{*0}\pi^0$ decays, the factorized amplitudes contain the form
factors, respectively, $F_0^{\pi B}(m_D^2)$ and 
$F_1^{\pi B}(m_{D^*}^{2})$, whose values are model dependent. However, 
we need not seriously worry about ambiguities arising from these form 
factors as long as the non-factorizable amplitudes which do not
involve them are dominant. [If the non-factorizable contribution is
not dominant, the results on these decays will be not very much
different from those of the usual factorization.] For the decay 
constants of heavy mesons, we assume $f_D \simeq f_{D^*}$ (and 
$f_B \simeq f_{B^*}$) since $D$ and $D^*$ ($B$ and $B^*$) are 
expected to be degenerate because of heavy quark symmetry\cite{HQET} 
and are approximately degenerate in reality. Here we take 
\newpage
\begin{center}
\begin{quote}
{Table~II. Factorized and non-factorizable amplitudes for the 
$\bar B \rightarrow D\pi$ and $D^*\pi$ decays.
The CKM matrix elements are factored out. }
\end{quote}
\vspace{0.5cm}

\begin{tabular}
{l|l|l}
\hline
$\quad\,\,${\rm Decay}
&$\quad A_{\rm fact}\,(\times 10^{-5}$  GeV)
&$\qquad\qquad A_{\rm non-f}\,(\times 10^{-5}$ GeV) 
\\
\hline
$\bar B^0 \rightarrow D^{+}\pi^-$
& $\quad 1.54\,a_1\Bigl\{
                  1 - 0.11\Bigl({a_2 \over a_1}\Bigr)\Bigr\}$
&$\, -3.52a_2B_H\,\Bigl\{
\Bigl[{4\over 3}e^{i\delta_1} -{1\over 3}e^{i\delta_3}\Bigr]
                                              - 0.55 \Bigr\}$
\\
$\bar B^0 \rightarrow D^{0}\pi^0$
& $\, -1.23\,a_2\Bigl\{
                         {f_D \over 0.205\,\,{\rm GeV}}\Bigr\}$
& $\, -3.52a_2B_H\,\Bigl\{
{\sqrt{2} \over 3}\Bigl[2e^{i\delta_1} + e^{i\delta_3}\Bigr]
                                              - 0.05 \Bigr\}$
\\
$B^- \rightarrow D^0\pi^-$
& $\quad 1.54\,a_1\Bigl\{
                   1 + 1.14\Bigl({a_2 \over a_1}\Bigr)\Bigr\}$ 
& $\quad 3.52a_2B_H\,\Bigl\{e^{i\delta_3} + 0.48 \Bigr\}$
\\
$\bar B^0 \rightarrow D^{*+}\pi^-$
& $\, -1.53\,a_1\Bigl\{
                    1 - 0.39\Bigl({a_2 \over a_1}\Bigr)\Bigr\}$
& $\quad 2.13a_2B_H\,\Bigl\{ \Bigl[
{4 \over 3}e^{i\delta_1^*} - {1 \over 3}e^{i\delta_3^*}
                                        \Bigr] - 0.91 \Bigr\}$
\\
$\bar B^0 \rightarrow D^{*0}\pi^0$
& $\quad 1.25\,a_2\Bigl\{{f_{D^*} \over 0.205\,\,{\rm GeV}}\Bigr\}$
& $\quad 2.13a_2B_H\,\Bigl\{
{\sqrt{2} \over 3}\Bigl[2e^{i\delta_1^*} + e^{i\delta_3^*}\Bigr]
                                                  + 0.02 \Bigr\}$
\\
$B^- \rightarrow D^{*0}\pi^-$
& $\, -1.53\,a_1\Bigl\{
                 1 + 1.20\Bigl({a_2 \over a_1}\Bigr)\Bigr\}$ 
& $\, -2.13a_2B_H\,\Bigl\{e^{i\delta_3^*} +0.95 \Bigr\}$
\\
\hline
\end{tabular}

\end{center}
\vspace{0.5cm}
$f_{D^*} \simeq f_D \simeq 205$ MeV and 
$f_{B^*} \simeq f_B \simeq 175$ MeV 
from a recent result of lattice QCD\cite{Lattice}, 
$f_D = 203 \pm 7 \pm 20$ MeV and $f_B = 178 \pm 9 \pm 18$  MeV. 
In this way, we can obtain the factorized amplitudes in the second 
column of Table~II, where we have neglected very small annihilation 
terms in 
the $\bar B^0 \rightarrow D^0\pi^0$ and 
$D^{*0}\pi^0$ decay amplitudes. 

To evaluate the non-factorizable amplitudes, we need to know the 
size of the asymptotic matrix elements of $H_w$ and $A_\pi$ taken 
between heavy meson states. The latter which was parameterized in 
Eq.(\ref{eq:MEA}) is estimated to be 
$|h|\simeq 1.0$\cite{HARDP,suppl} by using partially conserved 
axial-vector current (PCAC) and the observed rate\cite{PDG}, 
$\Gamma(\rho \rightarrow \pi\pi)_{expt} \simeq 150$ MeV. 
To estimate the asymptotic matrix elements, 
\begin{equation}
\langle D^0|H_w|\bar B^0 \rangle, \,\,
\langle D^{*0}|H_w|\bar B^0 \rangle,  \,\,
\langle D^0|H_w|\bar B^{*0} \rangle \quad{\rm and}\quad  
\langle D^{*0}|H_w|\bar B^{*0} \rangle,      \label{eq:AMHW}
\end{equation} 
included in the non-factorizable amplitudes, we apply the
factorization to them since the heavy mesons annihilate at the weak 
vertex in the weak boson mass $m_W \rightarrow \infty$ limit; 
for example, 
\begin{equation}
\langle D^0|H_w|\bar B^0 \rangle 
= {G_F\over \sqrt{2}}V_{cb}V_{ud}\Bigl({m_D^2+m_B^2\over 2}\Bigr)
         B_Hf_Df_Ba_2,                         \label{eq:fac-AMHW}
\end{equation} 
in the IMF, where $H_w=H_w^{BSW}$ and $B_H$ is an analogue to the $B$ 
parameter in the matrix element of $O_{\Delta B=2}$ providing the
$B-\bar B$ mixing. Since the other asymptotic matrix elements of 
$H_w$ in Eq.(\ref{eq:AMHW}) can be estimated in the same manner as 
in Eq.(\ref{eq:fac-AMHW}), we obtain the hard pion amplitudes as 
the non-factorizable contributions listed in the third column of 
Table~II where we have used 
$f_{D^*} \simeq f_D \simeq 205$ MeV and 
$f_{B^*} \simeq f_B \simeq 175$ MeV  
as before and used the same $B_H$ parameter for all the asymptotic
matrix elements in Eq.(\ref{eq:AMHW}). 
The CKM matrix elements have been factored out. 

We now estimate branching ratios, 
$B(\bar B\rightarrow D\pi)$ and $B(\bar B \rightarrow D^*\pi)$, 
by taking a sum of the factorized amplitude (the second column in 
the Table~II) and the non-factorizable amplitude (the third column in 
Table~II) as the total amplitude\cite{footnote}. To this, we have to 
give  values of remaining parameters. We take $U_{cb}=0.038$ from the 
updated value $|U_{cb}|=0.0388 \pm 0.0036$\cite{Ali}. For the 
coefficients $a_1$ and $a_2$ in the effective weak Hamiltonian 
$H_w (=H_w^{BSW})$, we do not know their true values. According to 
Ref.\cite{BURAS}, NLO corrections to $a_1$ are small while 
corresponding corrections to $a_2$ may be large (approximately of the 
same size as the LO corrections, which seems to be unnatural from the 
view point of the perturbation theory) and very unstable with respect 
to change of renormalization scheme as mentioned before. We expect 
that the value of $a_1$ with the LO corrections is not far from the 
true value and that higher order corrections should be small as is 
expected formally in the perturbation theory. Therefore we take 
conservatively $a_1=1.03$ and $a_2=0.11$ which are given in the LO 
approximation\cite{NRSX}. For the phases $\delta_1$ and $\delta_3$ 
($\delta_1^{*}$ and $\delta_3^{*}$) arising from contributions of 
non-resonant multi-hadron intermediate states with isospin 
$I={1\over 2}$ and ${3\over 2}$ which have not been measured by 
experiments, they will be restricted in the region 
$|\delta_{2I}|,\,|\delta_{2I}^{*}| < 90^\circ$. 
(Resonant contributions have already been extracted as pole 
amplitudes in $M_S$, although they were not very important and were 
neglected as discussed before.) We here assume 
$\delta_1 \simeq \delta_1^*$ and $\delta_3 \simeq \delta_3^*$ 
because of the heavy quark symmetry. For the value of the $B_H$ 
parameter, we here consider two cases, $B_H=1$ and $0.5$ as typical 
examples, and list the branching ratios for 
$U_{cb}=0.038$, $U_{ud}=0.98$, $\delta_1=\delta_1^*=60^\circ$, 
$\delta_3=\delta_3^*=-60^\circ$ 
in Table~III, where we have used the observed lifetime\cite{PDG}, 
$\tau(B^-) \simeq \tau(\bar B^0)\simeq 1.5\times10^{-12}$ s. 
$B_{\rm fact}$, $B_{\rm non-f}$ and $B_{\rm total}$ are estimated
branching ratios which include only the factorized amplitude, only
the non-factorizable one and the sum of them, respectively. 
It is seen that the non-factorizable contributions to the 
\newpage
\begin{quote}
{Table~III. Branching ratios ($\%$) for $\bar B \rightarrow D\pi$ and 
$D^*\pi$ decays where $a_1=1.03$, $a_2=0.11$, $U_{cb}=0.038$, 
$U_{ud}=0.98$ and $\tau(B^-) = \tau(\bar B^0) = 1.5\times10^{-12}$ s. 
The values $\delta_1=\delta_1^*=60^\circ$ and 
$\delta_3=\delta_3^*=-60^\circ$ of phases are taken tentatively. 
$B_{\rm fact}$, $B_{\rm non-f}$ and $B_{\rm total}$ include only the
factorized amplitude, only the non-factorizable one and the sum of
them, respectively.  
The data values are taken from Ref.\cite{BHP}.}

\end{quote}

\begin{center}
\begin{tabular}
{l|l|l|l|l}
\hline
{$\quad${\rm Decay}}
&$\quad B_{\rm fact}\quad $
&{\begin{tabular}{c|c}
\multicolumn{2}{c}{$\quad B_{\rm non-f}\quad $} \\
\hline
{ $B_H=1$ } & {$B_H=0.5$} 
\end{tabular}
}
&{\begin{tabular}{c|c}
\multicolumn{2}{c}{$\,\, B_{\rm total}\,\, $} \\
\hline
 { $B_H=1$ } & { $B_H=0.5$ }
\end{tabular}
}
&$\quad B_{\rm expt}\quad $\\
\hline
$\bar B^0 \rightarrow D^{+}\pi^-$
&{\quad 0.26}
&{\begin{tabular}{l|l}
{\,\quad  0.033\,\,\,} & {\,\,\,\,\, 0.008\quad }
\end{tabular}
}
&{\begin{tabular}{l|l}
{\,\quad 0.30\quad} & {\,\quad 0.27 \quad}
\end{tabular}
}
&$\,\,0.31 \pm 0.04$
\\
\hline
$\bar B^0 \rightarrow D^{0}\pi^0$
&{\quad 0.002\,\,\,}
&{\begin{tabular}{l|l}
{\,\quad 0.010\,\,\,} & {\,\quad 0.002} 
\end{tabular}
}
&{\begin{tabular}{l|l}
{\,\quad 0.019\,\,\,} & {\,\quad 0.008}
\end{tabular}
}
& $<0.048$
\\
\hline
$B^- \rightarrow D^0\pi^-$
&\quad 0.34
&{\begin{tabular}{l|l}
{\,\quad 0.027\,\,\,} & {\,\quad  0.007} 
\end{tabular}
}
&{\begin{tabular}{l|l}
 {\,\quad 0.51\quad} & {\,\quad 0.42}
\end{tabular}}
&$\,\,0.50 \pm 0.07$
\\
\hline
$\bar B^0 \rightarrow D^{*+}\pi^-$
&\quad 0.24
&{\begin{tabular}{l|l}
{\,\quad 0.013\,\,\,} & {\,\quad 0.003} 
\end{tabular}
}
&{\begin{tabular}{l|l}
 {\,\quad 0.29\quad} & {\,\quad 0.26}
\end{tabular}}
&$\,\,0.28 \pm 0.05$
\\
\hline
$\bar B^0 \rightarrow D^{*0}\pi^0$
&$\quad$0.002\,\,
&{\begin{tabular}{l|l}
{\,\quad 0.004\,\,\,} & {\,\,\,\,\, 0.001\quad} 
\end{tabular}
}
&{\begin{tabular}{l|l}
 {\,\quad 0.011\,\,\,} & {\,\quad 0.006}
\end{tabular}
}
&$<0.097$
\\
\hline
$B^- \rightarrow D^{*0}\pi^-$
&\quad 0.34
&{\begin{tabular}{l|l}
{\,\quad 0.024\,\,\,} & {\,\quad  0.006} 
\end{tabular}
}
&{\begin{tabular}{l|l}
{\,\quad 0.49\quad} & {\,\quad  0.41}
\end{tabular}
}
&$\,\,0.52 \pm 0.10$
\\
\hline
\end{tabular}

\end{center}
\vspace{0.5 cm}
color favored $\bar B \rightarrow D\pi$ and $D^*\pi$ decays are small 
but still can interfere efficiently with the main amplitude given by
the naive factorization and that the long distance hadron dynamics 
can improve remarkably 
$B(\bar B \rightarrow D\pi)_{\rm fact}$ and 
$B(\bar B \rightarrow D^*\pi)_{\rm fact}$. 
The predicted branching ratios 
$B(\bar B \rightarrow D\pi)_{\rm total}$ and 
$B(\bar B \rightarrow D^*\pi)_{\rm total}$ 
for the color favored $\bar B \rightarrow D\pi$ and $D^*\pi$ decays 
reproduce well the observed ones. 

In the color suppressed 
$\bar B^0 \rightarrow D^0\pi^0$ and $D^{*0}\pi^0$ decays, the
non-factorizable contributions are considerably larger than the 
factorized ones in the case of $B_H \simeq 1$ while the former is 
comparable with the latter in the case of $B_H \simeq 0.5$. 
Therefore, the ambiguities arising from the uncertainties of the 
values of $F_0^{\pi B}(m_D^2)$ and $F_1^{\pi B}(m_{D^*}^2)$ which are 
involved in the factorized amplitudes for these decays are not very 
serious as long as $B_H \simeq 1$. Within the present approximation,
the predicted branching ratio 
$B(\bar B^0 \rightarrow D^0\pi^0)_{total}$ will be not much lower 
than the present experimental upper limit if $B_H \simeq 1$ while the 
former will be much less than the latter if $B_H \lesssim 0.5$. 

\section{$\bar B \rightarrow J/\psi\bar K$ and $J/\psi\pi$ decays}

Now we study Cabibbo-angle favored $\bar B \rightarrow J/\psi\bar K$ 
and suppressed $B^- \rightarrow J/\psi\pi^-$ decays in the same way 
as in the previous section. Both of them are color suppressed and 
their kinematical condition is much different from the color favored 
$\bar B \rightarrow D\pi$ and $D^*\pi$ decays at the level of 
underlying quarks, {\it i.e.}, $b \rightarrow (c\bar c)_8\,+\,s$ in
the former but $b \rightarrow c\,+\,(\bar ud)_1$ in the latter. 
Therefore, dominance of factorized amplitudes in the 
$\bar B \rightarrow J/\psi\bar K$ and $B^- \rightarrow J/\psi\pi^-$ 
decays has no theoretical support and hence it is expected that 
non-factorizable long distance contribution is important in these 
decays. 

The factorized amplitude for the 
$\bar B \rightarrow J/\psi\bar K$ decays is given by 
\begin{equation}
M_{\rm fact}(\bar B \rightarrow J/\psi\bar K)
=-iU_{cb}U_{cs}
\Bigl\{
{G_F \over \sqrt{2}}a_2f_\psi F_1^{KB}(m_\psi^2)
\Bigr\}2m_\psi\epsilon^*(p')\cdot p. 
\end{equation}
The value of the decay constant of $J/\psi$ is estimated to be 
$f_\psi \simeq 0.38$ GeV from the observed rate\cite{PDG} for the 
$J/\psi \rightarrow \ell^+\ell^-$. The value of the CKM matrix 
element $U_{cs}$ is given by $U_{cs} \simeq U_{ud} \simeq 0.98$. 
The value of the form factor $F_1^{KB}(m_\psi^2)$ has not been 
measured and its theoretical estimates are model dependent. 
We pick out tentatively the values of $F_1^{KB}(m_\psi^2)$ based on 
the following five models, {\it i.e.}, BSW\cite{BSW}, GKP\cite{GKP}, 
CDDFGN\cite{CDDFGN}, AW\cite{AW} and ISGW\cite{ISGW}, and list the 
corresponding $B(\bar B \rightarrow J/\psi\bar K)_{\rm fact}$ in 
Table~IV, where we have used $U_{cb}=0.038$, 
$\tau_B = 1.5\times 10^{-12}$ s and $a_2=0.11$ as before. The results 
($B_{fact}$) from the factorized amplitudes for all the values of 
$F_1^{KB}(m_\psi^2)$ listed in Table~IV are much smaller 
than the observations\cite{BHP}, 
\begin{eqnarray}
&&B(B^- \rightarrow J/\psi K^-)_{\rm expt} 
= (0.102 \pm 0.014) \,\,\% \nonumber \\
&&B(\bar B^0 \,\rightarrow J/\psi\bar K^0\,)_{\rm expt} 
= (0.075 \pm 0.021) \,\,\%. \label{eq:psi-K}
\end{eqnarray}

Non-factorizable contributions to these decays are estimated by using 
a hard kaon approximation which is a simple extension of the hard 
pion technique in the previous section. With this approximation and
isospin symmetry, non-factorizable amplitude for the 
$\bar B \rightarrow J/\psi\bar K$ decays is given by 
\begin{equation}
M_{\rm non-f}(\bar B \rightarrow J/\psi\bar K) 
= {i \over f_K}\langle \psi|H_w|\bar B_s^0 \rangle
\Biggl\{e^{i\delta_{\psi\bar K}} 
+ {\langle \psi|H_w|\bar B_s^{*0} \rangle 
\over \langle \psi|H_w|\bar B_s^{0} \rangle }
{m_B^2 - m_\psi^2 \over m_{B_s^*}^2 - m_\psi^2}
\sqrt{1 \over 2}h 
\Biggr\}\, +\, \cdots, 
\end{equation}
\newpage 
\begin{quote}
{Table~IV. Branching ratios ($\%$) for the 
$\bar B \rightarrow J/\psi\bar K$ decays where the values of 
$F_1^{KB}(m_\psi^2)$ estimated in the five models, BSW, GKP, CDDFGN, 
AW and ISGW, in Refs.\cite{BSW}, \cite{GKP}, \cite{CDDFGN}, 
\cite{AW} and \cite{ISGW}, respectively, and tentatively chosen 
$|\delta_{\psi\bar K}|=60^\circ$ are used. Values of the other 
parameters involved are the same as in Table~III.  
The data values are taken from Ref.\cite{BHP}.}

\end{quote}

\begin{center}
\begin{tabular}
{l|l||c|c|c|c|c}
\hline
\multicolumn{2}{l||}
{Models}
& $\quad$BSW$\quad$
& $\quad$GKP$\quad$
& $\quad$CDDFGN$\quad$
& AW
& $\quad$ISGW$\quad$ 
\\
\hline
\multicolumn{2}{l||}
{$F_1^{K B}(m_\psi^2)$}
&$\quad$0.565$\quad$
&$\quad$0.837$\quad$
&$\quad$0.726$\quad$
&$\quad$0.542$\quad$
&$\quad$0.548$\quad$
\\
\hline
\multicolumn{2}{l||}
{$B_{\rm fact}$} 
&0.010
&0.022
&0.016
&0.009
&0.009 
\\
\hline
$B_{\rm non-f}$
&{
\begin{tabular}{l}
{$B_H'=1$}\\ 
{$B_H'=0.5$}
\end{tabular}
}
&\multicolumn{5}{c}
{
\begin{tabular} {c}
{0.057} \\
{0.014}
\end{tabular}
}
\\
\hline
$B_{\rm total}$
&{
\begin{tabular}{l}
{$B_H'=1$}\\
{$B_H'=0.5$}
\end{tabular}
}
&{\begin{tabular} {c}
{0.11} \\
{\,\,\,0.045}
\end{tabular}
}
&{\begin{tabular} {c}
{0.14}
\\
{\,\,\,0.067}
\end{tabular}
}
&{\begin{tabular} 
{c}
{0.13}\\
{\,\,\,0.058}
\end{tabular}
}
&{\begin{tabular} 
{c}
{0.11}\\
{\,\,\,0.044}
\end{tabular}
}
&{\begin{tabular} 
{c}
{0.11}\\
{\,\,\,0.044}
\end{tabular}
}
\\
\hline
\multicolumn{2}{l||}{Experiment}
&\multicolumn{5}{c}
{
\begin{tabular}{c}
$B(B^-\rightarrow J/\psi K^-) = (0.102 \pm 0.014)\,\,\%$ 
\\
$B(\bar B^0\, \rightarrow J/\psi\bar K^0\,) 
= (0.075 \pm 0.021)\,\,\%$
\end{tabular}
}
\\
\hline
\end{tabular}

\end{center}
\vspace{0.5 cm}
where $\delta_{\psi\bar K}$ is the phase from contributions 
of multi-hadron intermediate states into the $\psi\bar K$ final
state and the ellipsis denotes neglected contributions of
excited mesons\cite{Close}.  We have used 
$\langle \bar B_s^0|V_{K^+}|B^- \rangle = -1$ and 
$\sqrt{2}\langle \bar B_s^{*0}|A_{K^+}|B^- \rangle = -h$ 
which are flavor $SU_f(3)$ extensions of Eqs.(\ref{eq:MEV}) and 
(\ref{eq:MEA}). Values of asymptotic matrix elements, 
$\langle \psi|H_w|\bar B_s^{0} \rangle$, etc., can be estimated by
using the factorization as before. Then the total amplitude for the 
$\bar B \rightarrow J/\psi\bar K$ decays is approximately given by 
\begin{eqnarray}
M(\bar B &&\rightarrow J/\psi\bar K)_{\rm total} \nonumber\\
&&\simeq -iU_{cb}U_{cs}\bigl\{5.73F_1^{KB}(m_\psi^2)\,
+\,B_H'\bigl[3.82e^{i\delta_{\psi\bar K}}\,+\,5.11\bigr]
\bigr\}a_2\times10^{-5}\quad {\rm GeV}         \label{eq:amp-psi-k}
\end{eqnarray}
where $f_K \simeq 0.16$ GeV and $f_{B^*_s} \simeq f_{B_s} \simeq 202$ 
MeV from the updated lattice QCD result\cite{Lattice}, 
$f_{B_s} = 202 \pm 26$ MeV, have been taken. $B_H'$ is a $B$ 
parameter corresponding to $B_H$ in Eq.(\ref{eq:fac-AMHW}). 

{}From the above amplitude, we see that 
$|M_{fact}(\bar B \rightarrow J/\psi\bar K)|
\lesssim |M_{non-f}(\bar B \rightarrow J/\psi\bar K)|$ 
unless $B_H'\lesssim 0.5$. If $B_H' \simeq 1$, the non-factorizable 
contribution will be dominant and 
$B(\bar B\rightarrow J/\psi\bar K)_{\rm total}$ from the amplitude in 
Eq.(\ref{eq:amp-psi-k}) can reproduce the observed values in 
Eq.(\ref{eq:psi-K}) when reasonable values of $|\delta_{\psi\bar K}|$ 
are taken. [Since the $\psi\bar K$ state is exotic, 
$\delta_{\psi\bar K}$ will be not very far from $\delta_3^*$ in the 
$\bar B \rightarrow D^*\pi$ decays.] We list our results for 
$|\delta_{\psi\bar K}|=60^\circ$ in Table~IV as an example. It is 
seen that ambiguities arising from uncertainty of 
$F_1^{KB}(m_\psi^2)$ which is included in the factorized amplitude 
will be considerably diluted because of the non-factorizable 
contribution if $B_H \simeq 1$ . 

For the Cabibbo-angle suppressed $B^- \rightarrow J/\psi\pi^-$, 
the same technique and values of parameters as the above lead to 
\begin{eqnarray}
M(B^-&& \rightarrow J/\psi\pi^-)_{\rm total} \nonumber\\
&&\simeq -iU_{cb}U_{cd}
\bigl\{5.73F_1^{\pi B}(m_\psi^2)\,
+\,B_H'\bigl[3.79e^{i\delta_{\psi\pi}}\,+\,5.52\bigr]
\bigr\}a_2\times10^{-5}\quad {\rm GeV}.       \label{eq:amp-psi-pi}
\end{eqnarray}
Using 
$F_1^{\pi B}(m_\psi^2) \simeq F_1^{KB}(m_\psi^2)$ 
and 
$\delta_{\psi \pi}\simeq \delta_{\psi\bar K}$ 
expected from $SU_f(3)$ symmetry, we obtain 
\begin{equation}
B(B^- \rightarrow J/\psi\pi^-)_{\rm total}\simeq 
\Biggl|{U_{cd} \over U_{cs}}\Biggr|^2
B(B^- \rightarrow J/\psi K^-)_{\rm total}
\end{equation}
which is well satisfied by experiments \cite{CLEOII,CDF}.
{}From Eq.(\ref{eq:amp-psi-pi}), it is seen that the non-factorizable 
long distance contribution is again dominant in this decay when 
$B_H' \simeq 1$. $B(B^- \rightarrow J/\psi\pi^-)_{\rm total}$ 
from the amplitude Eq.(\ref{eq:amp-psi-pi}) which includes both of 
the factorized amplitude and the non-factorizable one can reproduce 
the existing experimental data, 
\begin{eqnarray}
B(B^- \rightarrow J/\psi\pi^-)_{\rm expt} 
 && = (4.7\,\pm\,2.6)\times 10^{-5}\quad 
                         ({\rm CLEO}\cite{CLEOII}), \nonumber\\
&& = (5.0\,\,\,^{+\,\,\,\,2.1}_{-\,\,\,\,1.9}\,\,)\times 10^{-5}
                  \quad ({\rm CDF}  \cite{CDF}),  
                                               \label{eq:psi-pi}
\end{eqnarray}
by taking $a_2\simeq 0.11$ and reasonable values of
$|\delta_{\psi\pi}|$ and $B_H'$.

\section{Summary}

In summary, we have investigated the $\bar B \rightarrow D\pi$,  
$D^*\pi$, $J/\psi \bar K$ and $J/\psi\pi^-$ decays describing the 
amplitude for these decays by a sum of factorizable and 
non-factorizable contributions. The former amplitude has been 
estimated by using the naive factorization while the latter has 
been calculated by using a hard pion (or kaon) approximation. 
The so-called final state interactions have been included in the 
non-factorizable long distance contributions. The non-factorizable 
contribution to the color favored $\bar B \rightarrow D\pi$ and 
$D^*\pi$ decays is rather small and therefore the final state 
interactions seem to be not very important in these decays although 
still not negligible. By taking reasonable values of the phase shifts 
arising from contributions of multi-hadron intermediate states to the 
non-factorizable amplitudes, the observed branching ratios for these 
decays can be well reproduced in terms of a sum of the hard pion 
amplitude and the factorized one when the values of the coefficients 
$a_1\simeq 1.03$ and $a_2\simeq 0.11$ of four quark operators in the 
BSW weak Hamiltonian are taken. Namely, the factorized amplitudes are 
dominant but not complete and long distance hadron dynamics should be 
carefully taken into account in hadronic weak interactions of $B$ 
mesons. If $a_2\simeq 0.2$, which may be given by the next-to-leading 
order corrections but very strongly dependent on the choice of 
renormalization scheme and unstable\cite{BURAS}, were taken, however, 
these decays might be saturated by the factorizable 
contributions\cite{CLEO,Kamal}, and then the sum of factorized and 
non-factorizable amplitudes would provide too large rates for these 
decays unless $B_H \ll 1$.  

In color suppressed $\bar B \rightarrow  D^0\pi^0$, $D^{*0}\pi^0$, 
$J/\psi \bar K$ and $J/\psi\pi^-$ decays, non-factorizable long
distance contributions are important. In particular, in the 
$\bar B^0 \rightarrow J/\psi\bar K^0$ decay, long distance physics 
should be treated carefully since it will play an important role to 
determine the $CP$ violating CKM matrix element. When 
$a_2\simeq 0.11$ are taken, a sum of factorized and non-factorizable 
amplitudes with reasonable values of the phases 
$|\delta_{\psi \bar K}|$ and $|\delta_{\psi\pi}|$ can reproduce the 
observed values of 
$B(\bar B \rightarrow J/\psi \bar K)$ and 
$B(B^- \rightarrow J/\psi\pi^-)$, 
respectively, although both of the present theory and the existing
data still contain large ambiguities. 

The non-factorizable amplitudes are proportional to asymptotic 
ground-state-meson matrix elements of $H_w$, {\it i.e.}, the $B$
parameter, $B_H$ or $B_H'$. If $B_H \simeq B_H'\simeq 1$, the
non-factorizable amplitudes will be dominant in the amplitudes for
the color suppressed decays and the value of 
$B(\bar B^0 \rightarrow D^0\pi^0)$ is predicted to be not much lower 
than the present experimental upper limit. 
If $B(\bar B^0 \rightarrow D^0\pi^0)$ is measured to be much less 
than the present upper limit, $B_H$ will be considerably smaller than 
unity. If $a_2\simeq 0.2$ were taken, the sum of the factorized and
non-factorizable amplitudes would give too large branching ratios for 
the color suppressed decays unless $B_H$ and $B_H'$ are much less 
than unity. 

Therefore more precize measurements of branching ratios for the color
suppressed decays, in particular, $B(\bar B \rightarrow D^0\pi^0)$ 
are useful to determine the non-factorizable long distance 
contributions in hadronic weak decays of $B$ mesons. 

\vskip 1cm

The author would like to thank to Prof. T.~E.~Browder and the other 
members of high energy physics group of University of Hawaii for 
their discussions, comments and hospitality during his stay there.


\begin{references}
\bibitem{footnote} The signs of the annihilation term in the
factorized amplitude and the surface term in the non-factorizable one 
are different from those in our previous work\cite{BDP} in which the 
phases of matrix elements of currents were chosen to reproduce the 
expression of the factorized amplitudes in Ref.\cite{Kamal}. 
However the choice of the phases of current matrix elements was 
confusing. Nevertheless, it did not lead to a serious problem since 
the contributions of the annihilation term in the factorized 
amplitude and the non-factorizable amplitude itself were rather small 
in color favored $\bar B\rightarrow D\pi$ and $D^*\pi$ decays.
\end{references}
\end{document}